\documentclass[a4paper,12pt]{article}

\newcommand{\sect}[1]{\setcounter{equation}{0}\section{#1}}

\begin{document}
\topmargin 0pt \oddsidemargin 0mm

\renewcommand{\thefootnote}{\fnsymbol{footnote}}
\begin{titlepage}
\begin{flushright}
 hep-th/0312014
\end{flushright}

\vspace{5mm}
\begin{center}
{\Large \bf Holography, the Cosmological Constant and \\
 the Upper Limit of the Number of e-foldings} \vspace{32mm}

{\large Rong-Gen Cai\footnote{e-mail address: cairg@itp.ac.cn}}

\vspace{10mm} {\em  Institute of Theoretical Physics, Chinese
Academy of Sciences,\\
 P.O. Box 2735, Beijing 100080, China }

\end{center}

\vspace{5mm} \centerline{{\bf{Abstract}}}
 \vspace{5mm}
If the source of the current accelerating expansion of the
universe is a positive  cosmological constant, Banks and Fischler
argued that there exists an upper limit of the total number of
e-foldings of inflation. We further elaborate on the upper limit
in the senses of viewing the cosmological horizon as the boundary
of a cavity and of the holographic D-bound in a de Sitter space.
Assuming a simple evolution model of inflation, we obtain an
expression of the upper limit in terms of the cosmological
constant, the energy density of inflaton when the inflation
starts, the energy density as the inflation ends, and reheating
temperature. We discuss how the upper limit is modified in the
different evolution models of the universe. The holographic
D-bound gives more high upper limit than the entropy threshold in
the cavity. For the most extremal case where the initial energy
density of inflation is as high as the Planck energy, and the
reheating temperature is as low as the energy scale of
nucleosynthesis, the D-bound gives the upper limit as $146$ and
the entropy threshold as $122$. For reasonable assumption in the
simplest cosmology, the holographic D-bound leads to a value about
$85$, while the cavity model gives a value around $65$ for the
upper limit, which is close to the value in order to solve the
flatness problem and horizon problem in the hot big bang
cosmology.

\end{titlepage}

\newpage
\renewcommand{\thefootnote}{\arabic{footnote}}
\setcounter{footnote}{0} \setcounter{page}{2}

\sect{Introduction}
 The holographic principle is perhaps one of most fundamental
 principles of nature~\cite{Holo}. This principle relates a theory including
 gravity in $D$ dimensions to a theory without gravity in lower
 dimensions. The AdS/CFT correspondence~\cite{AdS} is a beautiful example of
 the realization of the holographic principle in an anti-de Sitter
 (AdS) space.  Two years ago, Strominger~\cite{Stro} argued that a similar
 correspondence (dS/CFT correspondence) may also exist in a de Sitter space,
 which is a maximally symmetric
  curved space of positive constant curvature.

 Needless to say, the de Sitter space plays an important role in modern
 cosmology. In the inflation model the universe is in a quasi-de
 Sitter phase.  Furthermore, a lot of astronomical
 observations~\cite{Super1,Super2,WMAP} indicate
 that the expansion of our universe is now accelerating, rather
 than decelerating. A most simple explanation of the observed
 accelerating is that our universe has a positive cosmological
 constant with the value ($\Lambda_0 \sim (10^{-3}ev)^4$)~\cite{WMAP}.
 In that case, out universe
 will approach again to a de Sitter
 phase in the far future.

 The inflation model~\cite{Guth,KT,Liddle}, which says that our universe has
 undergone an accelerating epoch in the early time, is now a
 popular paradigm, in which some fundamental difficulties in the hot
 big bang cosmology can be solved, for example, the
 problems of the spatial flatness, the large-scale smoothness, the
 small-scale inhomogeneity, and the unwanted relics, etc..
A remarkable series of observations support the inflation model
(for example, see \cite{WMAP2}). It is then natural to ask whether
the holography principle has any consequence to the inflation
model. In Ref.~\cite{AKS} Albrecht, Kaloper and Song have argued
that the holography could lead to a nontrivial energy scale (UV
cutoff) above which the description of the effective field theory
for perturbations during the inflation losses its validness.
However, it results in a series of controversies~\cite{Contro}:
the D-bound could not give us a nontrivial UV cutoff; instead it
is satisfied naturally. The so-called D-bound~\cite{Bousso} says
that there is a maximal entropy bound of a system in a de Sitter
space; the bound is given by the difference between the
cosmological horizon areas of pure de Sitter space and
asymptotically de Sitter space containing the system.

On the other hand, it is well known that in order to solve
 the spatial flatness problem and the horizon puzzle,
there is a low limit of the number of e-foldings of
inflation~\cite{Guth,KT}. Quite interestingly it has been shown
robustly that there is also an upper limit for the number ($\tilde
N$) of e-foldings before the end of inflation at which observable
perturbations were generated~\cite{Hui,LL}. However, it says
nothing about the total number ($N$) of e-foldings of inflation,
which is usually a much large number than $\tilde N$. If the
current acceleration  is due to a cosmological constant, very
recently Banks and Fischler~\cite{BF} have argued that there
exists an upper limit of the total number ($N$) of e-foldings of
inflation, $e^N= a_e/a_i$. Here $a_e$ and $a_i$ denote the scale
factors of the universe at the end of inflation and at the
beginning of inflation, respectively. Obviously both the numbers
($\tilde N$ and $N$) of e-foldings of inflation play a crucial
role in building a successful inflation model, which is still
lacking, and in understanding the evolution of the universe. In
the brane world scenario, a similar upper limit has been discussed
in Ref.~\cite{Wang}.

In this note we would like to discuss further the upper limit of
the total number of e-foldings of inflation. In the next section,
we first review the Banks-Fischler's limit briefly. With a simple
assumption of evolution of the universe, we then obtain an
expression of the upper limit in terms of the cosmological
constant, the initial energy density and end energy density during
inflation, and the reheating temperature,  and discuss some
subtleties of the upper limit. In Sec.~3 we use the D-bound to
give the upper limit of the total number of e-foldings and discuss
some related points. The conclusion is included in Sec.~4.

\sect{Banks-Fischler's Limit}

The key point of inflation to solve the  spatial flatness problem
and the horizon puzzle is the production of a big amount of
entropy during reheating after the end of inflation~\cite{Guth}.
The starting point to get the upper limit of the total number of
e-foldings is also the entropy in the universe.

 For a fluid with an
equation of state, $p=\kappa \rho$, contained in a finite cavity
with radius $R$,  the authors of Ref.~\cite{Fischler} have shown
that there is an upper bound of entropy stored in the fluid,
beyond which black holes will form. The upper bound has the
scaling relation to the radius
\begin{equation}
\label{2eq1}
 S_R \sim R^{3-2/(1+\kappa)}.
 \end{equation}
 This relation holds in four dimensions\footnote{In $(n+1)$-dimensions
 we find that the upper bound of entropy should have the form $S_R \sim
 R^{n-(n-1)/(1+\kappa)}$.}. On the other hand, a static observer in a de Sitter
 space with cosmological constant, $\Lambda_0$, cannot see the
 whole space, the boundary of the region which can be causally accessible to the
 observer is called the cosmological horizon with size
 $R_c \sim 1/\sqrt{\Lambda_0}$. Banks and Fischler~\cite{BF} argued that the
 cosmological horizon could be the natural boundary of the cavity
 with radius $R_c$. Then the threshold of entropy of the fluid that is accessible to
 the static observer in the de Sitter space is\footnote{In some
 expressions concerning entropy throughout the paper, a numerical factor of order ${\cal
 O}(1)$ has been omitted. Such a factor has a negligible effect to the
 upper limit since it appears as $\ln {\cal O}(1)$ in the expressions of
  the upper limit. See also \cite{BF}.}
 \begin{equation}
 \label{2eq2}
 S_C \sim \Lambda_0^{-\frac{1 +3\kappa}{2(1+\kappa)}}.
 \end{equation}
 It is interesting to note that in order to avoid the big crush
 for a closed Friedmann-Robertson-Walker (FRW) universe with a
 positive cosmological constant, the upper bound of the entropy of the
  fluid in
 the universe is also given by (\ref{2eq2}) \cite{Fischler}. Of
 course, it should be reminded that for a flat or open FRW
 universe with a positive cosmological constant, one cannot reach
 this conclusion.

 Now let us consider thermodynamics of the universe. The Friedmann equation describing
 the evolution of a flat FRW  universe with a positive cosmological constant, $\Lambda$,
 is
 \begin{equation}
 \label{2eq3}
 H^2 =  \rho +\frac{\Lambda}{3}.
 \end{equation}
 Using the
 thermodynamic relation, that relates the energy density to the
 entropy density of a thermodynamic system,
 \begin{equation}
 \label{2eq4}
 \sigma \sim \rho^{1/(1+\kappa)},
 \end{equation}
 one can write the entropy of the fluid in the FRW
  universe with scale factor $a$ as
 \begin{equation}
 \label{2eq5}
 S \sim a^3 \rho^{1/(1+\kappa)}.
 \end{equation}
 Consider the moment at which the inflation ends, and denote the energy
 density of the universe
 by $\rho_e$. Suppose that the initial size of the causal patch is given by
 the Hubble radius at the beginning of inflation, $H^{-1}_i$. During
 inflation this patch grows by a factor of $e^N$; the scale factor at the end
 of inflation is then $a \sim
 \exp(N) H^{-1}_i$.  Using (\ref{2eq5}), the total
 entropy in that patch therefore is
 \begin{equation}
 \label{2eq6}
 S \sim \exp (3N) H^{-3}_i \rho_e^{1/(1+\kappa)}.
 \end{equation}
 Having considered that the change of the Hubble radius is small during
  inflation, one can take
  \begin{equation}
  \label{2eq7}
  \rho_e \sim H^2_i\sim \Lambda_I,
  \end{equation}
  where $\Lambda_I$ is the value of the energy density during
  inflation. The entropy (\ref{2eq6}) of fluid  should be less than the one
  (\ref{2eq2}), which leads to the Banks-Fischler's limit of the
  total number of e-foldings~\cite{BF}
  \begin{equation}
  \label{2eq8}
  N < \frac{1+3\kappa}{6(1+\kappa)}\ln
  \frac{\Lambda_I}{\Lambda_0}.
  \end{equation}
  It should be stressed that the universe has been assumed to be
  asymptotically de Sitter here.
  Note that the dark energy density  has the same order as the critical
  density,  so  $\Lambda_0 \sim
  (10^{-3}ev)^4$. Suppose that the energy density of
  inflation is at $\Lambda_I \sim (10^{16}Gev)^4$, one then has
  \begin{equation}
  \label{2eq9}
  N < 85,
  \end{equation}
  in the case of $\kappa=1$, for the stiff matter. Such an
  equation of state appears when the universe is filled
  with a kinetic energy dominated scalar field. On the other hand, one has
  \begin{equation}
  \label{2eq10}
  N < 65,
  \end{equation}
  in the case of $\kappa =1/3$, for the radiation matter. It is interesting to note
  that these
  values for the total number of e-foldings of inflation are close
  to the value necessary to solve the horizon problem and flatness
  problem~\cite{BF,Guth}. However, we find that in the
  matter dominated phase  with $\kappa=0$, for the $\Lambda_I$ and
  $\Lambda_0$ given above,
  \begin{equation}
  \label{2eq11}
  N < 43.
  \end{equation}
  Note that in the usual inflation model, the inflation is
  followed by a matter dominated phase where the inflaton is
  rapidly oscillating about a minimum of its potential~\cite{KT,Liddle}.
  The value (\ref{2eq11}) seems too small and is seemingly excluded
  according to the current astronomical observations~\cite{Hui,LL}.

  Of course the above estimation is very rough. In this section we
  are interested in what the maximal and minimal upper limits of the total
  number of e-foldings are within possible inflation models, if one
  insists that the entropy threshold
  in the asymptotically de Sitter space is given by (\ref{2eq2}).
  For this end, it seems worth to recall the main evolution
  processes of the universe~\cite{KT}, in particular, the processes before
  the radiation dominated phase.  In a simple inflation
  model, for example, the single field model, suppose that the inflation starts
  with the size $H_i^{-1}$ of
  the causal patch and  the
  energy density $\rho_i$, in which the potential of the inflaton dominates,
  and ends with energy density $\rho_e$. During inflation, the
  causal path expands to $\exp (N) H^{-1}_i$. After inflation, the
  universe goes into a coherent oscillation phase, in which
  the inflaton is rapidly oscillating around a minimum of its potential. After some
  time, the inflaton  decays rapidly to particles in the standard model and the universe
  is ``reheated".  Following the reheating is just the ordinary adiabatic, radiation-dominated
  phase of the hot big bang cosmology. During the reheating, the universe grows by a
  factor of $a_{re}/a_{e} \sim (\rho_e/T^4_{re})^{1/3}$, where $\rho_e$ is the energy density
   at the end of inflation and $T_{re}$ denotes the reheating
  temperature, which is the initial temperature of the radiation
  epoch. When the energy density of radiation matter equals to
  that of nonrelativistic matter, the universe enters into the matter
  dominated phase.  As is well known, the total
  entropy within a casual patch is conserved in the standard
  cosmology model. In other words, the current total entropy mainly from the
  radiation matter of
  our observable universe is in the same order as that during the
  radiation dominated  phase.

  With the above consideration, the total entropy of our universe
  at the beginning of the radiation dominated phase (at that moment the radiation
  temperature is just the reheating temperature), is\footnote{It
  is the entropy of radiation matter. Note that the number of particle species
  is cancelled here.}
  \begin{equation}
  \label{2eq12}
  S \sim  \exp(3N) H_i^{-3} (\rho_e/T^4_{re}) T^3_{re}.
  \end{equation}
In this case, the entropy bound $S < S_C$ yields
\begin{equation}
\label{2eq13}
 N < \frac{1}{4} \ln \frac{\rho_i}{\Lambda_0}
    +\ln \frac{\rho_i^{1/4}}{T_{re}} +\frac{4}{3}\ln
    \frac{T_{re}}{\rho_e^{1/4}}.
\end{equation}
Comparing (\ref{2eq8}) (take $\kappa=1/3$) with our result
(\ref{2eq13}), it is easy to see that they give the same value
only when $\rho_i \sim \rho_e \sim T^4_{re}$. Further we note that
the right hand side of (\ref{2eq13}) can be rewritten as
\begin{equation}
\label{2eq14}
N < \frac{1}{4} \ln \frac{\rho_i}{\Lambda_0}
  +\frac{1}{4}\ln \frac{\rho_i}{\rho_e} +\frac{1}{12}\ln
  \frac{T^4_{re}}{\rho_e}.
\end{equation}
Obviously it is not an easy matter to determine the upper limit of
the number of e-foldings of inflation since we are still not yet
very clear for many events during evolution of the universe, in
particular, for events which happened before nucleosynthesis. In
other words, there are a lot of uncertainties to determine the
upper limit.  So following \cite{LL}, we now consider how the
upper limits of the number of e-foldings is modified as one
changes the properties of inflation model within the range allowed
by theories and observations.

(1) {\it The extremal case.} In the chaotic inflation
model~\cite{Linde}, the initial energy density could be as high as
the Planck scale, $\rho_i \sim (10^{19}Gev)^4$. On the other hand,
the most extremal assumption could be that the reheating continues
almost to the nucleosynthesis, which happens at the energy scale
$(10^{-3}Gev)^4$, though the electroweak scale $(10^2Gev)^4$ is
regarded usually as the practical limit at the end of
inflation~\cite{LL}. In addition, in supersymmetric theories in
order to avoid the overproduction of gravitinos, the energy
density should be below $(10^{11}Gev)^4$. Due to the red shift of
energy during reheating, the reheating temperature is usually less
than the energy density at the end of inflation (that is, the
third term of
 the right hand side of equation (\ref{2eq14}) gives a negative
contribution to the total number), so the extremal assumption is
that the most high reheating temperature is of the order of
$\rho_e^{1/4}$. Consider the three energy scales given above as
$\rho_e$, we find from (\ref{2eq14}) that
\begin{eqnarray}
\label{2eq15}
 && N \sim  122, ~~~{\rm for}\  \ \rho_e \sim (10^{-3}Gev )^4;
 \ \ \ N \sim 110,~~~{\rm for}\ \ \rho_e \sim
 (10^2Gev)^4;\nonumber \\
 &&~~~~~~~~~~~~~~~~~~~~~~~~~~~~~~~~~~~~~  {\rm and}\ N \sim 90,
  ~~~{\rm for}\ \ \rho_e \sim
 (10^{11}Gev)^4,
\end{eqnarray}
respectively. Although the differences among the three energy
scales are large, the differences of total numbers of e-foldings
are not so large as expected.

(2) {\it A plausible upper limit.} A plausible energy scale at
which the inflation starts is the GUT scale with $\rho_i \sim
(10^{16} Gev)^4$. It is well known that during inflation, the
potential goes down very slowly. Suppose that the inflation ends
at $\rho_e \sim (10^{14}Gev)^4$. Although there are still
considerable uncertainties for the reheating temperature, the
total number of e-foldings is not very sensitive to the third term
on the right hand side of equation (\ref{2eq14}) since there is a
suppressing factor of $(1/12)$ in this term.  For example, we have
 \begin{eqnarray}
 \label{2eq16}
 && N \sim 68, ~~~{\rm for}\ T_{re} \sim 10^{12}Gev; ~~~
    N \sim 62, ~~~~{\rm for}\ T_{re} \sim 10 ^5 Gev; \nonumber \\
    && ~~~~~~~~~~~~~~~~~~~~~~~~~~ {\rm and} \ \ N \sim 56, ~~~{\rm
    for}\ T_{re} \sim 10^{-3}Gev,
    \end{eqnarray}
    respectively. With the given $\rho_i$ and $T_{re}$,
    a more high energy density $\rho_e$ at the end of inflation will
    lead to a lower upper limits than those given by (\ref{2eq16}).
     An inflation model with more low energy scale is
    that the inflation begins with $\rho_i \sim (10^{14}Gev)^4$,
    which satisfies the constraint of gravity wave amplitude from
    the observation of CMB anisotropy~\cite{Hui}. Suppose the
    inflation ends with $\rho_e \sim (10^{12} Gev)^4$, we then
    have
    \begin{eqnarray}
    \label{2eq17}
    && N \sim 65, ~~~{\rm for}\ T_{re} \sim 10^{12}Gev; ~~~
    N \sim 61, ~~~~{\rm for}\ T_{re} \sim 10 ^8 Gev; \nonumber \\
    && ~~~~~~~~~~~~~~~~~~~~~~~~~~ {\rm and} \ \ N \sim 58, ~~~{\rm
    for}\ T_{re} \sim 1Tev.
    \end{eqnarray}
    These values are obviously close to the minimal value
    necessary to solve the spatial flatness problem and horizon
    problem~\cite{Guth,KT}.

    (3) {\it The model of $\lambda \phi^4$.} This model is special
    in the sense that reheating in this model has an unusual
    feature. Usually the universe is in a matter dominated phase
    during the scalar field oscillation. For the $\lambda
    \phi^4$ model, however, the expansion of the universe is as
    radiation dominated~\cite{Ford}. In this case, the duration of the epoch
    of reheating no longer matters and one can take the universe
    as radiation dominated beginning at the end of
    inflation~\cite{LL}. As a result, the factor $\rho_e/T^4_{re}$
    in (\ref{2eq12}) is absent. Then we have
    \begin{equation}
    \label{2eq18}
    N < \frac{1}{4} \ln \frac{\rho_i}{\Lambda_0} +\frac{1}{4} \ln
    \frac{\rho_i}{T^4_{re}}.
    \end{equation}
    During the coherent oscillation phase in the usual inflation models, much
    of the initial vacuum energy is red shifted away, by a factor
    of $\rho_e/T^4_{re}$. In general $T^4_{re}$ therefore is always less
    than $\rho_e$. In the $\lambda \phi^4$ model, due to the absence of the factor
    $\rho_e/T^4_{re}$, we have the conclusion that within the same
    other conditions (the same initial potential of inflation and the same reheating
    temperature), the upper limit of the number of e-foldings
    in the $\lambda \phi^4$ model is larger than the usual
    inflation models, for example, the scalar field model with
    a quadratic potential.

 (4) {\it The stiff matter case.}  In the literature there are
      proposals to end inflation by making the inflation field
      a transition from a potential dominated phase to a kinetic
      energy dominated phase (see \cite{LL} and references therein).
       The equation of state of a  kinetic
      dominated inflation field is the one of stiff matter with
      $\kappa=1$. In this era, the universe expands with $a \sim
      t^{1/3}$ and $\rho \sim 1/a^6$. During this epoch, the
      universe grows by a factor $(\rho_e/T^4_{re})^{1/6}$. Here
      we have assumed that the stiff matter dominates from the end
      of inflation to the beginning of radiation dominated phase in the
      big bang cosmology. In this case, the expression
      (\ref{2eq12}) should be replaced by
      \begin{equation}
      \label{2eq19}
      S \sim \exp(3N) H_i^{-3}(\rho_e/T^4_{re})^{1/2} T^3_{re}.
      \end{equation}
      As a result, we get the upper limit of the total number of
      e-foldings
      \begin{equation}
      \label{2eq20}
      N < \frac{1}{4}\ln \frac{\rho_i}{\Lambda_0} +\frac{1}{4} \ln
      \frac{\rho_i}{\rho_e} +\frac{1}{12}\ln
      \frac{\rho_e}{T^4_{re}}.
      \end{equation}
      Because $\rho_e > T^4_{re}$, comparing (\ref{2eq20}) with (\ref{2eq14}),
      we see that the upper limit is raised in this evolution
      model of the universe.  In contrast to the simplest cosmology
      described
      above, the upper limit in this stiff matter case increases by $1/6 \ln
      (\rho_e/T^4_{re})$. This value is not large, compared to the first
      two terms on the side of right hand of (\ref{2eq20}). For example, it
      is only $1.53$ if $\rho_e \sim (10^{16}Gev)^4$ and $T_{re} \sim
      10^{12}Gev$.

\sect{Holographic Limit}
 Bekenstein~\cite{Bek} was the first to consider the issue of maximal entropy
 for a macroscopic system contained in a given region. For a
 closed system with total energy $E$, which fits in a sphere with
 radius $R$ in three spatial dimensions, He argued that there
 exists an upper bound on the entropy of the system, $S_B \le 2\pi
 E R$. This bound is often referred to as the Bekenstein bound.
 This bound is believed to be valid for a system with negligible
 self-gravity in a flat spacetime. In an asymptotically de Sitter
 space, based on the fact that the cosmological horizon of any
 asymptotically de Sitter spaces is always less than the one of a
 pure de Sitter space, Bousso~\cite{Bousso} argued that the maximal entropy of a
 system embedded in a de Sitter space is limited by
 \begin{equation}
 \label{3eq1}
 S_D = \frac{1}{4G} (A_0-A),
 \end{equation}
 by applying the generalized second law~\cite{Bek2} of black hole thermodynamics
 to the cosmological horizon. Here $A_0$ denotes the cosmological
 horizon area of the pure de Sitter space, while $A$ for the
 asymptotically de Sitter space. This bound (\ref{3eq1}) is often
 called the D-bound. Although the D-bound is the counterpart of the Bekenstein
 bound in de Sitter space~\cite{Bousso2,Cai}, it is applicable to a
 strongly self-gravity system, even including black holes, in de
 Sitter space.

 As shown in the previous section, the Banks-Fischler's limit
 is obtained on the basis that there exists an entropy threshold for a fluid
 in a cavity, beyond which black holes will form in the cavity; and that
 the de Sitter horizon is viewed as the boundary of the
 cavity. On the one hand, the de Sitter space is certainly
 different from the cavity. On the other hand, considerable series
 of evidence indicate that a lot of black  holes exist in our
 universe. As a result the assumption of Banks and Fischler may be
 too stringent. In this section we use the D-bound to give an
 upper limit of total number of e-foldings of inflation.

 Applying the D-bound to the current universe, one has
 \begin{equation}
 \label{3eq2}
 S_D  \sim  (\Lambda_0^{-1}-H^{-2}_0)
      \sim  \frac{0.3}{\Lambda_0}.
     \end{equation}
 Here  $\Lambda_0$ is the cosmological constant as before, $H_0$ is the
 current Hubble constant, and the $0.3$ is obtained by the current
 observations: $\Omega_M \sim 0.3$ and $\Omega_{\Lambda}\sim 0.7$.
 If one uses the D-bound at the beginning of the radiation
 dominated phase, the D-bound gives
 \begin{equation}
 \label{3eq3}
 S_D \sim \Lambda_0^{-1},
 \end{equation}
 since at that moment $H^{-2} \ll \Lambda_0^{-1}$. Therefore during
 evolution of the universe, the D-bound gives us the almost same
 entropy threshold of matter in the universe.

 Within the simple evolution process of inflation described in the previous section,
  the entropy (\ref{2eq12}) of radiation matter is limited by
 the D-bound (\ref{3eq3}), which leads  to
 \begin{equation}
 \label{3eq4}
 N <\frac{1}{3}\ln \frac{\rho_i}{\Lambda_0}
    +\frac{1}{6}\ln \frac{\rho_i}{\rho_e}
    +\frac{1}{6}\ln \frac{m^2_{pl}T^2_{re}}{\rho_e},
 \end{equation}
 where $m_{pl} \sim 10^{19}Gev$ is the Planck mass. Note that the
 main contribution to the upper limit comes from the term
 $\ln(\rho_i/\Lambda_0)$. Therefore, in general the holographic limit
 (\ref{3eq4}) is larger than the Banks-Fischler limit. For
 instance, in the extremal case, namely the case (1) discussed in
 the previous section, we have
 \begin{eqnarray}
\label{3eq5}
 && N \sim  146, ~~~{\rm for}\  \ \rho_e \sim (10^{-3}Gev )^4;
 \ \ \ N \sim 134,~~~{\rm for}\ \ \rho_e \sim
 (10^2Gev)^4;\nonumber \\
 &&~~~~~~~~~~~~~~~~~~~~~~~~~~~~~~~~~~~~~  {\rm and}\ N \sim 114,
  ~~~{\rm for}\ \ \rho_e \sim
 (10^{11}Gev)^4.
\end{eqnarray}
Further, for comparison, we calculate the upper limits in the case
of $\rho_i \sim (10^{16}Gev)^4$ and $\rho_e \sim (10^{14}Gev)^4$.
They are
\begin{eqnarray}
 \label{3eq6}
 && N \sim 91, ~~~{\rm for}\ T_{re} \sim 10^{12}Gev; ~~~
    N \sim 86, ~~~~{\rm for}\ T_{re} \sim 10 ^5 Gev; \nonumber \\
    && ~~~~~~~~~~~~~~~~~~~~~~~~~~ {\rm and} \ \ N \sim 80, ~~~{\rm
    for}\ T_{re} \sim 10^{-3}Gev,
    \end{eqnarray}
    respectively. Obviously, as the case of Banks-Fischler's
    limit, the holographic limit also depends on properties of inflation
    models and evolution process of the universe.

\sect{Conclusion}
 The current dark energy scale is $\rho \sim (10^{-3}ev)^4$,
 which is much low than the inflation energy scale, which is
 generally believed to be around the GUT scale $\sim(
 10^{16}Gev)^4$. Therefore people might think that the dark
 energy, which enforces the universe to accelerating expand now,
 has nothing to do with the inflation happened at the very
 early time. However, if the accelerating expansion is attributed to
 a positive cosmological constant, it does have something to do
 with the inflation model and with the fate of the universe. Banks and
 Fischler argued that there exists an upper limit of the total number of
 e-foldings of inflation in terms of the
 energy scale of inflation and the cosmological constant $\Lambda_0$,
 if one views the universe
 as a cavity with radius $1/\sqrt{\Lambda_0}$. Although there are
 a lot of uncertainties to determine the upper limit definitely, we further
 elaborate on and detail the upper limit: within a simple assumption of
 evolution of the universe, we obtain an expression of the
 upper limit of the total number of e-foldings of inflation in
 terms of the cosmological constant, the initial energy density
 and end energy density during inflation, and the reheating
 temperature. In the extremal case, where inflation is assumed to
 happen at the Planck scale and the reheating continues to the
 energy scale of nucleosynthesis, the cavity model gives us a value about
 $122$ for the upper limit, while the D-bound leads to a more
 large value $146$. Within a reasonable assumption for energy scales
 of inflation,
 the upper limit is around $65$, which is close to the value
 necessary to solve the spatial flatness problem and horizon
 problem in the hot big bang cosmology. We also discuss how the
 upper limit is modified when one changes the properties of
 inflation models.

\section*{Acknowledgments}
The author would like to thank Miao Li, Bin Wang, Dehai Zhang and
Xinmin Zhang for useful discussions. This work was supported in
part by a grant from Chinese Academy of Sciences, a grant from
NSFC, a grant from the Ministry of Education of China, and by the
Ministry of Science and Technology of China under grant No.
TG1999075401.

\end{document}